\documentclass{ifacconf}

\usepackage{graphicx}      % include this line if your document contains figures
\usepackage{natbib}        % required for bibliography
\usepackage{subcaption}
\usepackage{amsmath}
\usepackage{amsfonts}
\usepackage{color} 
\usepackage{algorithm}
\usepackage[]{algpseudocode}
\usepackage{xcolor}
\usepackage{booktabs}
\begin{document}
\begin{frontmatter}

\title{Generative design of stabilizing controllers with diffusion models: the Youla approach\thanksref{footnoteinfo}} 

\thanks[footnoteinfo]{This paper is partially supported by FAIR (Future Artificial Intelligence Research) project, funded by the NextGenerationEU program within the PNRR-PE-AI scheme (M4C2, Investment 1.3, Line on Artificial Intelligence), by the Italian Ministry of Enterprises and Made in Italy in the framework of the project 4DDS (4D Drone Swarms) under grant no. F/310097/01-04/X56 and by the PRIN PNRR project  P2022NB77E “A data-driven cooperative framework for the management of distributed energy and water resources” (CUP: D53D23016100001), funded by the NextGeneration EU program. It has been also partly supported by the project “European Lighthouse to Manifest Trustworthy and Green AI” (ENFIELD), which have received funding from the European Union’s Horizon Europe research and innovation program under grant agreement No. 101120657.}

\author[PoliMI]{Matteo Cercola}, 
\author[UM]{Donatello Materassi}, 
\author[PoliMI]{Simone Formentin}

\address[PoliMI]{Dipartimento di Elettronica, Informazione e Bioingegneria, Politecnico di Milano, Milano, Italy.(e-mail: name.surname@polimi.it).}

\address[UM]{Department of Electrical and Computer Engineering, University of Minnesota, Twin Cities. (mater013@umn.edu).}

\begin{abstract}
Designing controllers that simultaneously achieve strong performance and provable closed-loop stability remains a central challenge in control engineering. This work introduces a diffusion-based generative framework for linear controller synthesis grounded in the Youla–Kučera parameterization, enabling the construction of stabilizing controllers by design. The diffusion model learns a conditional mapping from plant dynamics and desired performance metrics to feasible Youla parameters, guaranteeing internal stability while flexibly accommodating user-specified targets. Trained on synthetically generated stable SISO plants with fixed-order Youla parameters, the proposed approach reliably synthesizes controllers that meet prescribed sensitivity and settling-time specifications on previously unseen systems. To the best of our knowledge, this work provides the first demonstration that diffusion models can generate stabilizing controllers, combining rigorous control-theoretic guarantees with the versatility of modern generative modeling.
\end{abstract}


\end{frontmatter}
%===============================================================================
\section{Introduction}

The synthesis of linear controllers that simultaneously meet multiple and potentially competing closed-loop performance requirements remains a longstanding challenge in control engineering \citep{bhattacharyya2022linear,zhou1998essentials,francis1987course}. Even for low-order linear time-invariant (LTI) systems, the relationship between controller parameters and performance indices, such as transient specifications or robustness margins, is highly nonlinear and difficult to characterize analytically \citep{khalil1996robust,skogestad2005multivariable}. As a consequence, controller design procedures often rely on indirect tuning rules \citep{borase2021review}, iterative numerical optimization \citep{gahinet1994linear}, or extensive parameter sweeps, all of which become computationally demanding when repeated closed-loop simulations are required. These challenges motivate the exploration of alternative paradigms for controller synthesis.

Recent advances in generative artificial intelligence, and in particular diffusion-based generative models (DMs) \citep{ho2020denoising,song2021scorebased}, offer a promising new direction. DMs excel at learning complex high-dimensional distributions from relatively modest datasets \citep{karras2022elucidating} and have demonstrated strong conditional generation capabilities even when the data lie on thin or low-dimensional manifolds \citep{de2022riemannian,oko2023diffusion}. 

The above property looks appealing for controller synthesis, as \textit{stabilizing controllers form a structured and low-measure subset of the full design space} \citep{Youla1976Optimal}. Furthermore, recent theoretical results show that DMs can efficiently and consistently learn such distributions \citep{lee2023convergence}, suggesting that they may be used to generate stabilizing controllers with user-specified performance characteristics. Motivated by these observations, this work proposes a diffusion-based framework for controller synthesis grounded in the Youla–Kučera parameterization. The objective is to generate \textit{fixed-order linear controllers} whose closed-loop behavior satisfies a set of performance metrics. Training data are constructed by randomly sampling stabilizing controllers in the Youla domain \citep{Mahtout2020Advances}, guaranteeing internal stability \textit{by construction}. As a case study, we condition the diffusion model on two closed-loop performance metrics, the settling time and the $\mathcal{H}_\infty$ norm of the sensitivity function \citep{zhou1998essentials}, so that it learns to produce Youla parameters that meet user-specified target values.

Restricting the conditioning to two classical metrics is intentional. Settling time and sensitivity capture complementary aspects of transient behavior and robustness \citep{skogestad2005multivariable}, while keeping the conditioning space interpretable and computationally manageable. This choice reduces dataset complexity, facilitates visualization, and isolates the core question: \emph{can a diffusion model learn the nonlinear mapping from Youla parameters to closed-loop specifications?} Our results indicate that it can. Despite its simplicity, the proposed conditional DM reliably generates stabilizing controllers that match the desired performance targets.

Experiments on a large set of simulated single-input single-output (SISO) LTI plants demonstrate that the model produces fixed-order controllers with high probability of stabilization and a substantial success rate in meeting the specified settling-time and sensitivity targets. These results provide initial evidence that diffusion-based generative modeling can serve as a viable and sample-efficient alternative to classical search or optimization-based controller design methods.

To the best of our knowledge, this work represents \textit{the first application of diffusion generative models to linear control design}. The proposed framework opens several promising research directions, including extensions to higher-order controllers, multi-input multi-output (MIMO) systems, richer performance specifications, constraints on control effort, and more structured sampling strategies in the Youla domain \citep{Mahtout2020Advances}. Together, these directions suggest that generative modeling may become a powerful tool for bridging modern machine learning with classical control design.

The remainder of the paper is as follows. In Section~\ref{sec:problem_formulation}, we formalize the controller-synthesis problem and introduce the Youla--Kučera representation adopted throughout the paper. Section~\ref{sec:dpm} provides a concise overview of diffusion probabilistic models and the noise-prediction training objective. Section~\ref{sec:main_contribution} details the proposed framework, including dataset construction, conditioning mechanisms, and the classifier-free guided sampling strategy. Section~\ref{sec:experimental_results} presents extensive numerical experiments on unseen plants, discussing success rates, failure modes, and qualitative behaviors of the generated controllers. A concluding section summarizes the main contributions and outlines a few interesting future research directions.

%%%%%%%%%%%%%%%%%%%%%%%%%%%%%%%%%%%%%%%%%%%%%%%%%%%%%%%%%%%%%%%%%

\section{Problem Formulation}\label{sec:problem_formulation}

The goal of this work is to synthesize stabilizing controllers for LTI systems by leveraging diffusion probabilistic models. We seek a data-driven framework that:
\begin{itemize}
    \item guarantees internal stability of the resulting closed-loop system,
    \item enables efficient and scalable dataset generation across diverse plant dynamics,
    \item produces controllers that satisfy prescribed closed-loop performance specifications.
\end{itemize}

Consider an LTI plant described by its transfer function $G(s)$. Instead of generating a controller $C(s)$ directly, we exploit the Youla--Kučera parameterization, which expresses every internally stabilizing controller as:
\begin{equation}\label{eq:youla}
    C(s) = \frac{Q(s)}{1 - G(s) Q(s)},
\end{equation}
where $Q(s) \in \mathcal{RH}_{\infty}$ is a stable and proper Youla parameter. This reformulation shifts the synthesis task to generating feasible Youla parameters. Since any stable, proper $Q(s)$ induces an internally stabilizing controller through the Youla--Kučera formula, closed-loop stability is guaranteed by construction once $Q(s) \in \mathcal{RH}_{\infty}$.

For each plant $G(s)$, we compute a vector of closed-loop performance metrics 
\[
    J = \big[ J_1, \ldots, J_K \big],
\]
representing the observed performance under a given controller. These metrics will later serve as conditioning variables for the generative model, enabling the synthesis of controllers that meet user-specified performance targets.

Our final objective is to learn a \textit{conditional generative model} that maps
\begin{equation}
    (G,\, J) \;\longmapsto\; Q,
\end{equation}
such that the Youla parameters $Q(s)$ sampled from the model induce controllers $C(s)$ that are both internally stabilizing and consistent with the desired closed-loop specifications encoded in $J$.

%%%%%%%%%%%%%%%%%%%%%%%%%%%%%%%%%%%%%%%%%%%%%%%%%%%%%%%%%%%%

\section{Denoising diffusion probabilistic models}\label{sec:dpm}

Denoising diffusion probabilistic models (DDPMs) \citep{sohldickstein2015deepunsupervisedlearningusing,ho2020denoising} are a class of generative models that construct data samples by reversing a gradual corruption process. Although originating from machine learning, their operating principle can be understood through concepts familiar to control engineers: the model learns the dynamics of a stochastic process that progressively removes noise from a signal, eventually recovering a clean sample drawn from the desired distribution.

\subsection{Forward diffusion: progressively adding noise}

Let $x_0$ denote a clean data vector. The \textit{forward diffusion process} generates increasingly noisy versions $\{x_i\}_{i=1}^N$ by iteratively injecting Gaussian noise:
\begin{equation}
    q(x_i \mid x_{i-1})
    = \mathcal{N}\!\left(\sqrt{1-\beta_i}\, x_{i-1},\, \beta_i I\right),
\end{equation}
where $\beta_i$ is a small variance increment. After many steps, $x_N$ becomes indistinguishable from a standard Gaussian vector. Because the dynamics are linear–Gaussian, the marginal distribution has the closed form
\begin{equation}
    q(x_i \mid x_0)
    = \mathcal{N}\!\left(
        \sqrt{\bar{\alpha}_i}\, x_0,\;
        (1-\bar{\alpha}_i) I
    \right),\qquad
    \bar{\alpha}_i = \prod_{j=1}^i (1-\beta_j).
\end{equation}

We stress here that the forward process is \textit{fixed} and no learning occurs here. It simply defines a stochastic “diffusion” phenomenon that gradually destroys the structure of the initial data.

\subsection{Reverse diffusion: learning to remove noise}

The core idea of DDPMs is to learn the reverse of this process: starting from pure noise $x_N \sim \mathcal{N}(0,I)$, the model reconstructs a clean sample $x_0$ by iteratively denoising. The reverse dynamics are modeled as
\begin{equation}
    p_\theta(x_{i-1} \mid x_i)
    = \mathcal{N}\!\bigl(x_{i-1} \mid \mu_\theta(x_i,i),\, \Sigma_i\bigr),
\end{equation}
where the covariances $\Sigma_i$ are fixed and the means $\mu_\theta(\cdot)$ are generated by a neural network with parameters $\theta$\footnote{From a control perspective, the reverse diffusion model plays a role analogous to that of an inverse-dynamics filter or observer used to reconstruct an uncorrupted input from a noisy or distorted measurement. The model learns the optimal denoising dynamics that undo the effect of the forward diffusion process, in the same way that classical filtering or channel equalization compensates for known corruption dynamics.}.

A key insight from \citep{ho2020denoising} is that learning the denoising policy is equivalent to learning the noise injected at each forward step. Using the reparameterization
\begin{equation}
    x_i = \sqrt{\bar{\alpha}_i}\, x_0 + \sqrt{1-\bar{\alpha}_i}\, \varepsilon,
    \qquad \varepsilon \sim \mathcal{N}(0,I),
\end{equation}
the neural network $\varepsilon_\theta(x_i,i)$ is trained to predict $\varepsilon$. The training objective is therefore
\begin{equation}
    \mathcal{L}(\theta)
    = \mathbb{E}_{x_0,\varepsilon,i}
      \left[\|\varepsilon - \varepsilon_\theta(x_i,i)\|^2\right],
\end{equation}
which corresponds to denoising score matching and serves as a tractable surrogate for maximizing data likelihood.

After training, sampling proceeds by initializing $x_N$ as random Gaussian noise and iteratively applying the learned reverse transitions $p_\theta(x_{i-1} \mid x_i)$. This process gradually removes noise and reconstructs a clean sample drawn from the learned data distribution.

%%%%%%%%%%%%%%%%%%%%%%%%%%%%%%%%%%%%%%%%%%%%%%%%%%%%%%%%%%%%%%

\section{Main Contribution}\label{sec:main_contribution}

\subsection{Dataset construction via Youla–based sampling}

Diffusion models require large amounts of representative training data. In domains such as computer vision, this is straightforward because vast image datasets already exist. In control, however, no comparable repositories of plant–controller pairs are available. As a result, the dataset for training a generative controller-synthesis model must be constructed from first principles.

In our framework, this can be made possible, \textit{e.g.}, by the Youla--Kučera parameterization \eqref{eq:youla}, which provides a complete and theoretically grounded description of all internally stabilizing controllers for a given plant. In fact,
we can generate stabilizing plant–controller pairs simply by sampling stable plants and feasible Youla parameters, \textit{without solving any optimization problem}. This offers a systematic and scalable mechanism for building the dataset.

For each sampled pair $(G, Q)$, we compute the associated controller $C$ and evaluate a set of closed-loop performance metrics
\[
\mathbf{J} = \big[J^{(1)}, \ldots, J^{(n_J)}\big].
\]
Samples that lead to physically unrealistic closed-loop behaviors are discarded. The remaining metrics are normalized through a $\log(1+J)$ transform, followed by standardization and removal of outliers outside a $3\sigma$ band. Plant and Youla coefficients are likewise standardized to ensure comparable numerical scales across all features.

Each dataset entry consists of:
\[
\mathrm{cond} = [\mathrm{coeff}(G),\, \mathbf{J}], 
\qquad 
x = \mathrm{coeff}(Q),
\]
where the conditioning vector encodes both the plant dynamics and the observed performance, while the target vector contains the coefficients of the corresponding Youla parameter. We adopt a fixed-dimensional coefficient-based representation for both $G(s)$ and $Q(s)$, avoiding the variability inherent in pole–zero or state-space formats.

Algorithm~\ref{alg:dataset_generation} summarizes the full procedure. The pipeline is computationally inexpensive and naturally parallelizable, making it compatible with the large-scale datasets that may be required by generative models.

\begin{algorithm}[t]
\caption{Dataset generation via Youla--Kučera parameterization}
\label{alg:dataset_generation}
\begin{algorithmic}[1]
\State Initialize dataset $\mathcal{D} \gets \emptyset$
\For{$k = 1$ to $N_{\text{samples}}$}
    \State Sample stable SISO plant $G_k(s)$
    \State Sample stable proper Youla parameter $Q_k(s) \in \mathcal{RH}_\infty$
    \State Compute controller:
    \[
        C_k(s) = \frac{Q_k(s)}{1 - G_k(s)Q_k(s)}
    \]
    \State Evaluate closed-loop metrics:
    \[
        \mathbf{J}_k = \big(J^{(1)}(G_k,C_k), \ldots, J^{(n_J)}(G_k,C_k)\big)
    \]
    \If{performance constraints are violated}
        \State \textbf{continue} \Comment{discard sample}
    \EndIf
    \State Normalize and standardize metrics and coefficients
    \State Conditioning vector: $\mathrm{cond}_k = [\mathrm{coeff}(G_k),\, \mathbf{J}_k]$
    \State Input vector: $x_k = \mathrm{coeff}(Q_k)$
    \State $\mathcal{D} \gets \mathcal{D} \cup \{(\mathrm{cond}_k, x_k)\}$
\EndFor
\State \Return $\mathcal{D}$
\end{algorithmic}
\end{algorithm}

\subsection{Training the diffusion model}
\label{sec:training_diffusion}

Once the dataset is constructed, we train a diffusion model to learn the conditional distribution of feasible Youla parameters. Recall that $x = \mathrm{coeff}(Q)$ denotes the vector of Youla coefficients. The forward diffusion process progressively corrupts $x$ with Gaussian noise, leading to a sequence of noisy variables $\{x_t\}$. The reverse process, parameterized by a neural network, learns how to remove noise step by step.

A key advantage of the Youla--Kučera parameterization is that every $Q$ in the training set corresponds to an internally stabilizing controller. Therefore, sampling from the learned distribution $p_\theta(x)$ guarantees internal stability \emph{by construction}. What remains is to steer the generative process toward meeting user-specified closed-loop performance targets.

Different strategies can be employed to incorporate performance specifications into the sampling procedure. The first two operate entirely at sampling time, assuming a diffusion model trained on the unconditional distribution $p(x)$ of feasible Youla parameters. The third strategy instead incorporates conditioning information into the reverse diffusion dynamics, thereby learning an approximation of the conditional distribution $p(x \mid \mathrm{cond})$.

\begin{itemize}
\item {\bf Guided sampling.}
A differentiable performance surrogate (a “classifier”) evaluates whether an intermediate sample meets the target. Its gradient nudges the diffusion trajectory \citep{dhariwal2021diffusionmodelsbeatgans,janner2022planningdiffusionflexiblebehavior}. This requires differentiability of $J$ with respect to $Q$ and increases computational cost.
\item {\bf Projection sampling.}
At each diffusion step, the noisy sample is projected onto the feasible set (e.g., stable $Q$, bounded metrics) via a small nonlinear optimization 
\citep{chung2024diffusionposteriorsamplinggeneral,zirvi2025diffusionstateguidedprojectedgradient}.
This avoids differentiating through $J$, but requires solving an optimization problem at every iteration.
\item {\bf Classifier-free guidance.}
In this approach, introduced in \citep{ho2022classifierfreediffusionguidance}, the same neural network is trained both conditionally and unconditionally, by randomly dropping the conditioning input during training. At sampling time, conditional and unconditional predictions are blended:
\begin{equation}
\varepsilon_t^{u} = \varepsilon_\theta(x_t, \emptyset, t), 
\qquad
\varepsilon_t^{c} = \varepsilon_\theta(x_t, \mathrm{cond}, t),
\end{equation}
\begin{equation}
    \hat{\varepsilon}_t = \varepsilon_t^{u} + \lambda (\varepsilon_t^{c} - \varepsilon_t^{u}),
\end{equation}
where $\lambda$ controls how strongly the generated controller is pushed toward the conditioning targets. This method is simple, efficient, and does not require differentiable performance metrics.
\end{itemize}

In this work, we will adopt the latter strategy, which is particularly attractive in control applications for a number of reasons:
\begin{itemize}
    \item it avoids differentiating closed-loop performance metrics, which are often nonsmooth or nonconvex;
    \item it requires no additional classifiers or optimization loops;
    \item it provides a lightweight mechanism for steering generation toward feasible high-performance controllers.
\end{itemize}

\begin{algorithm}[t]
\caption{Sampling the generative model for controller synthesis}
\label{alg:sampling}
\begin{algorithmic}[1]
\Require $\mathrm{cond}$: conditioning information (plant $G$ and desired metrics $\mathbf{J}$)
\Require $\lambda$: guidance strength
\State $\mathbf{x}_T \sim \mathcal{N}(0,\mathbf{I})$
\For{$t = T, \ldots, 1$}
    \State $\mathbf{z} \sim \mathcal{N}(0,\mathbf{I})$ \textbf{if} $t>1$ \textbf{else} $\mathbf{z}=0$
    \State $\varepsilon_t^{u} = \varepsilon_\theta(\mathbf{x}_t, \emptyset, t)$
    \State $\varepsilon_t^{c} = \varepsilon_\theta(\mathbf{x}_t, \mathrm{cond}, t)$
    \State $\hat{\varepsilon}_t = \varepsilon_t^{u} + \lambda (\varepsilon_t^{c} - \varepsilon_t^{u})$
    \State $\mathbf{x}_{t-1} =
        \frac{1}{\sqrt{\alpha_t}}
        \left(
            \mathbf{x}_t -
            \frac{1-\alpha_t}{\sqrt{1-\bar{\alpha}_t}}
            \hat{\varepsilon}_t
        \right)
        + \sigma_t \mathbf{z}$
\EndFor
\State \Return $\mathbf{x}_0$
\end{algorithmic}
\end{algorithm}

\subsection{Formal guarantees on controllers generation}

Recent theoretical results strengthen the suitability of diffusion models for structured domains such as stabilizing controller synthesis. In particular, \cite{oko2023diffusion} show that diffusion probabilistic models trained via denoising score matching are \emph{minimax-optimal} estimators for broad classes of smooth distributions. 

In our construction, the data distribution arises from smooth transformations of bounded Youla parameters and plant coefficients, with unstable or degenerate samples removed. This results in a compact, smooth, and well-behaved distribution that \textit{satisfies the regularity conditions} assumed in \cite{oko2023diffusion}. Consequently, the learned diffusion model \textit{inherits the same optimality and convergence guarantees}, providing theoretical support for approximating the distribution of stabilizing controllers achieving specified performance levels.

%%%%%%%%%%%%%%%%%%%%%%%%%%%%%%%%%%%%%%%%%%%%%%%%%%%%%%%%%%%%%%%%%%
\section{Experimental Results}\label{sec:experimental_results}

This section evaluates the proposed diffusion-based framework on a large collection of unseen SISO LTI plants. All experiments use second-order stable plants, and the closed-loop performance is assessed in terms of the $\mathcal{H}_\infty$ norm of the sensitivity function $\|S(s)\|_{\infty}$ and the settling time. The training set contains $200{,}000$ stabilizing plant--controller pairs generated via Algorithm~\ref{alg:dataset_generation}. The state vector is defined as $x=coeff(Q)$, comprising six parameters: three numerator coefficients and three denominator coefficients of the Youla parameter $Q(s)$.
The conditioning vector includes a total of eight parameters, six describing the plant $G(s)$ and two corresponding to the closed-loop performance metrics.

\begin{figure}[h!]
    \centering
    \includegraphics[width=\linewidth]{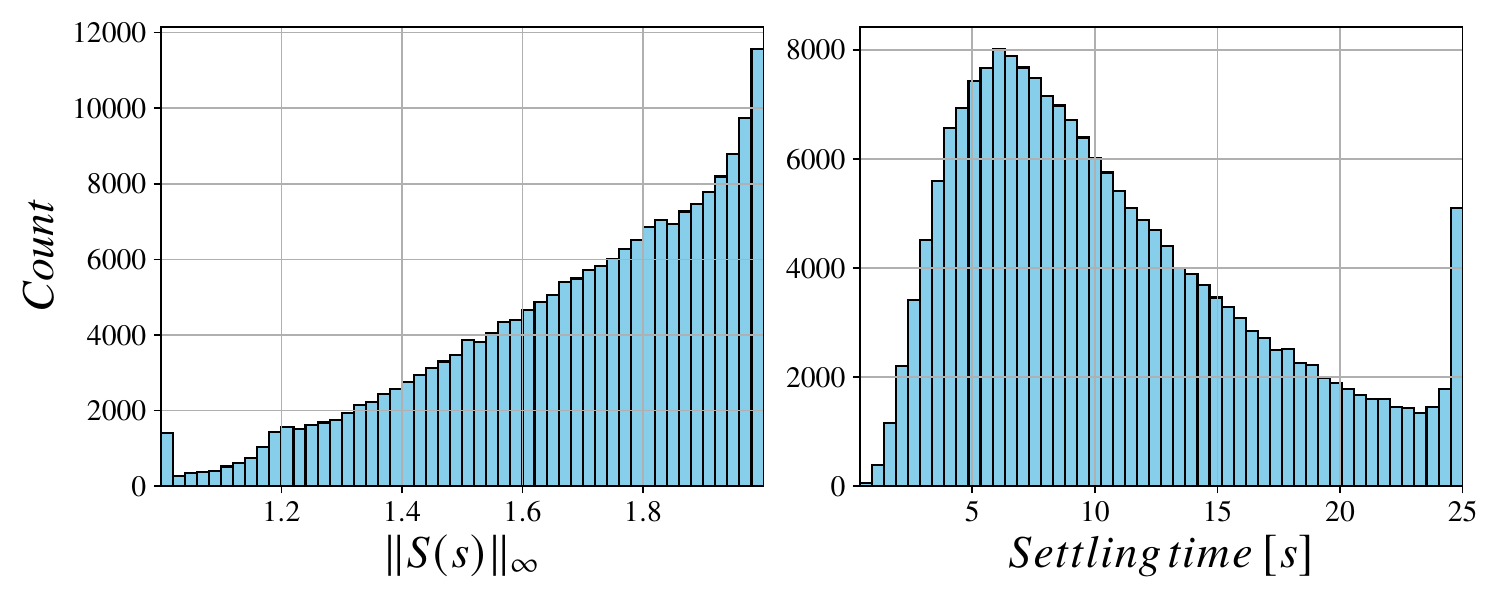}
    \caption{Distribution of closed-loop performance metrics in the training dataset ($200{,}000$ samples).}
    \label{fig:dataset_J_dist}
\end{figure}

Figure~\ref{fig:dataset_J_dist} shows the empirical distribution of the training metrics. To avoid extreme or noninformative behaviors, samples with $\|S(s)\|_{\infty}>2$ or settling time above $20\,\mathrm{s}$ were discarded. The remaining dataset spans a wide operational region, which is essential for learning a robust conditional generative model.

During testing, the diffusion model receives an unseen plant $G(s)$ together with a performance target vector $\mathbf{J}$ and generates $15$ candidate controllers through repeated sampling of the reverse diffusion process. This multishot generation is a key advantage of diffusion models, enabling rapid exploration of the performance–stability landscape for each plant.

\begin{figure*}[t]
    \centering
    \begin{subfigure}[b]{0.48\textwidth}
        \centering
        \includegraphics[width=\linewidth]{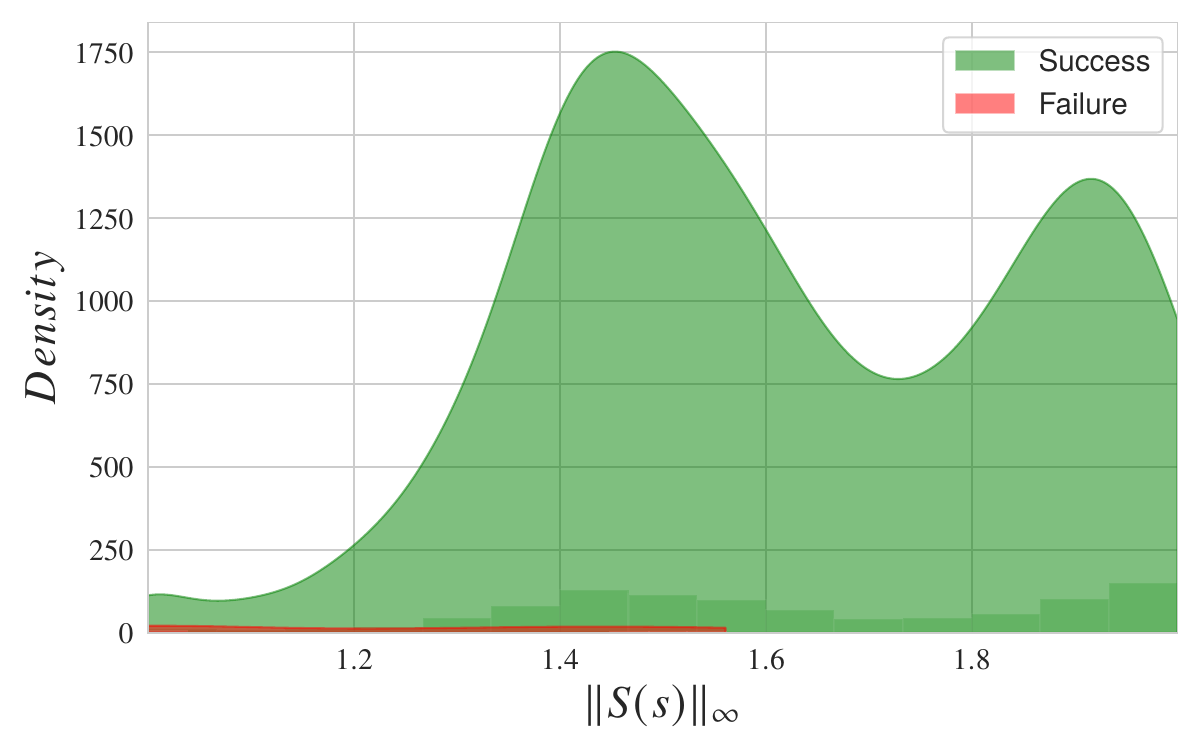}
        \caption{$\|S(s)\|_{\infty}$ distribution.}
    \end{subfigure}
    \hfill
    \begin{subfigure}[b]{0.48\textwidth}
        \centering
        \includegraphics[width=\linewidth]{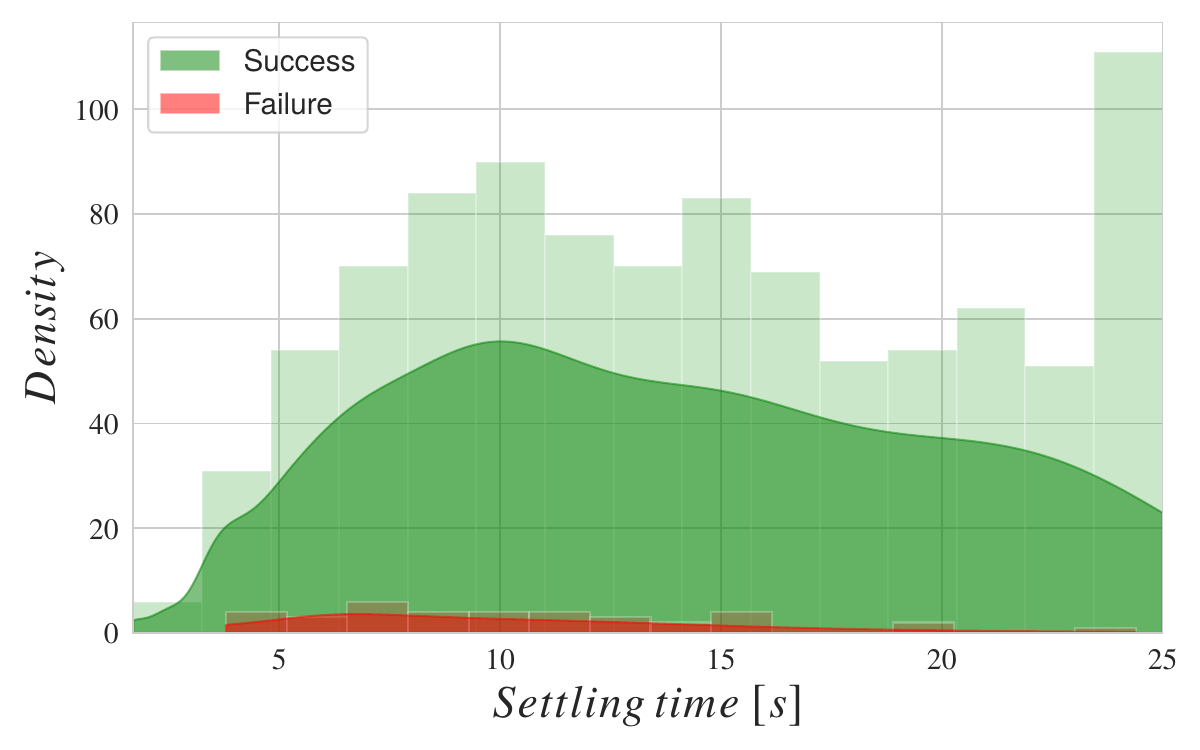}
        \caption{Settling time distribution.}
    \end{subfigure}
    \caption{Distribution of success and failure cases over $1{,}000$ unseen plants. A failure occurs when at least one target $J_i$ falls below $\widehat{J}_i - 2\sigma_i$.}
    \label{fig:kde_dist}
\end{figure*}

A sensitivity analysis over $200$ unseen plants was performed to select the \textit{guidance strength}~$\lambda$.  
Table~\ref{tab:lambda_success} reports the success rates, showing that $\lambda=1.1$ provides the best overall performance.

\begin{table}[h]
\centering
\begin{tabular}{lccccc}
\toprule
$\lambda$      & 0.5 & 1.0 & 1.1 & 2.0 & 4.0 \\
\midrule
Success (\%) & 94\% & 95\% & \textbf{95.5\%} & 92\% & 90\% \\
\bottomrule
\end{tabular}
\caption{Success rates for different values of the \textit{guidance strength} $\lambda$ on 200 unseen plants.}
\label{tab:lambda_success}
\end{table}

We finally evaluate the trained model on a test set of $1{,}000$ unseen plants. A generation is considered \emph{successful} if all requested targets $J_i$ satisfy
\begin{equation}
    J_i \ge \widehat{J}_i - 2\sigma_i,
\end{equation}
where $\widehat{J}_i$ and $\sigma_i$ are the mean and standard deviation of the $i$-th metric across the 15 generated controllers. This criterion is conservative: it allows the generated controllers to outperform the target but flags cases where the requested metric lies significantly outside the model's achievable region.

Across all $1{,}000$ test cases, \textit{the model succeeds in {95.3\%} of the plants}.  
Figure~\ref{fig:kde_dist} shows the distribution of successful and failed generations as a function of the target metrics. Success cases dominate the density for both sensitivity and settling time, confirming that the diffusion model generalizes well across unseen plants.

\begin{figure}[h!]
    \centering
    \includegraphics[width=\linewidth]{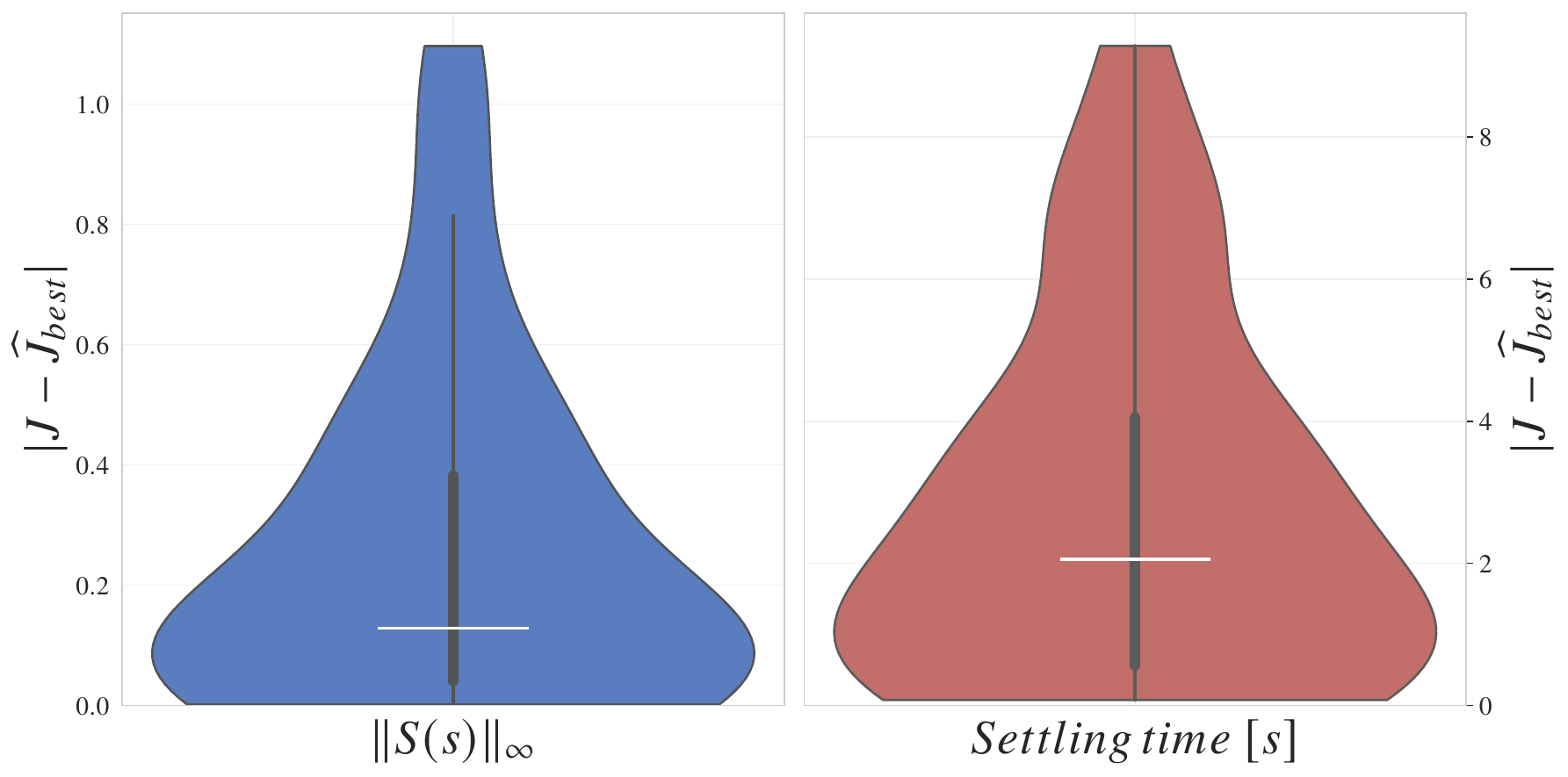}
    \caption{Absolute deviation between the target and the best generated controller across the 4.7\% failure cases.}
    \label{fig:failure_violin_plot}
\end{figure}

To understand the model's performance when the requested target is unrealizable or overly aggressive, we measure the deviation
\[
|J_i - \widehat{J}_{i,\mathrm{best}}|,
\]
where $\widehat{J}_{i,\mathrm{best}}$ is the performance of the best among the 15 generated controllers.  
Figure~\ref{fig:failure_violin_plot} shows that deviations remain small (median $0.13$ in $\|S(s)\|_{\infty}$ and $2\,\mathrm{s}$ in settling time), indicating that even when requirements cannot be fully met, the model produces controllers lying close to the achievable performance frontier.

\begin{figure}[h!]
    \centering
    \includegraphics[width=\linewidth]{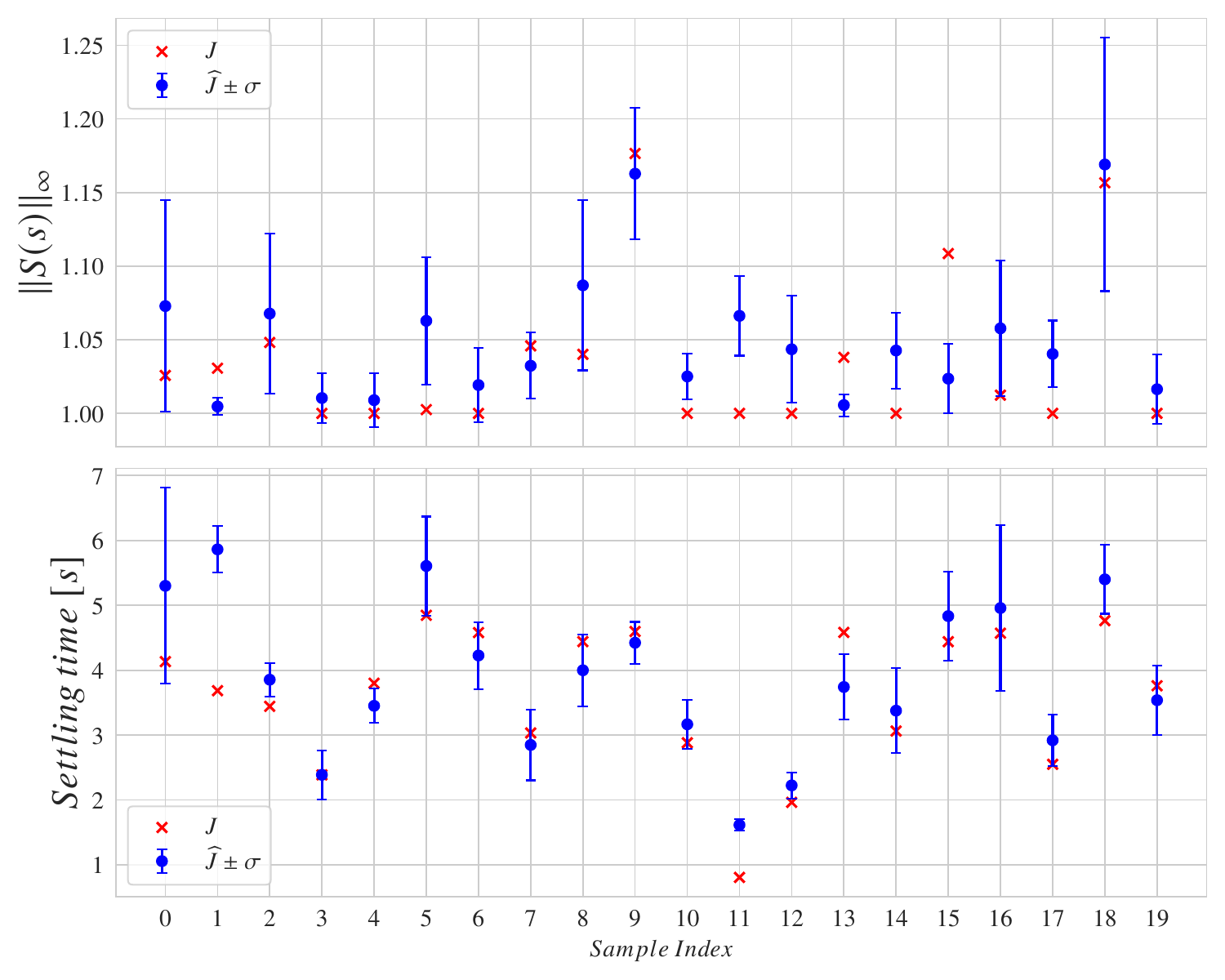}
    \caption{Generation results for $20$ high-performance test cases. Red: desired metrics; blue: mean and standard deviation of the generated controllers.}
    \label{fig:samples_jhat}
\end{figure}

\begin{figure*}[t]
    \centering
    \begin{subfigure}[b]{0.48\textwidth}
        \centering
        \includegraphics[width=\linewidth]{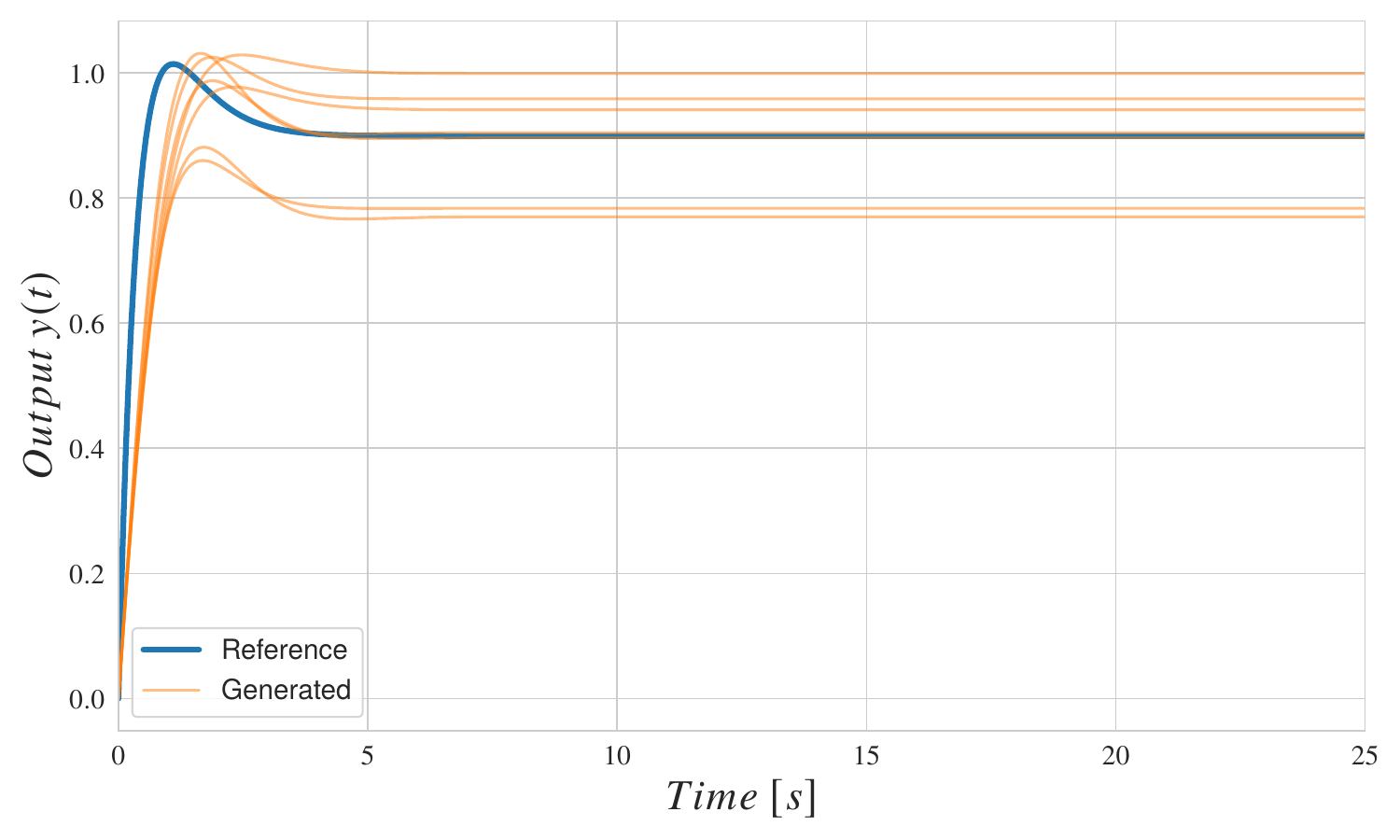}
        \caption{Input step response.}
    \end{subfigure}
    \hfill
    \begin{subfigure}[b]{0.48\textwidth}
        \centering
        \includegraphics[width=\linewidth]{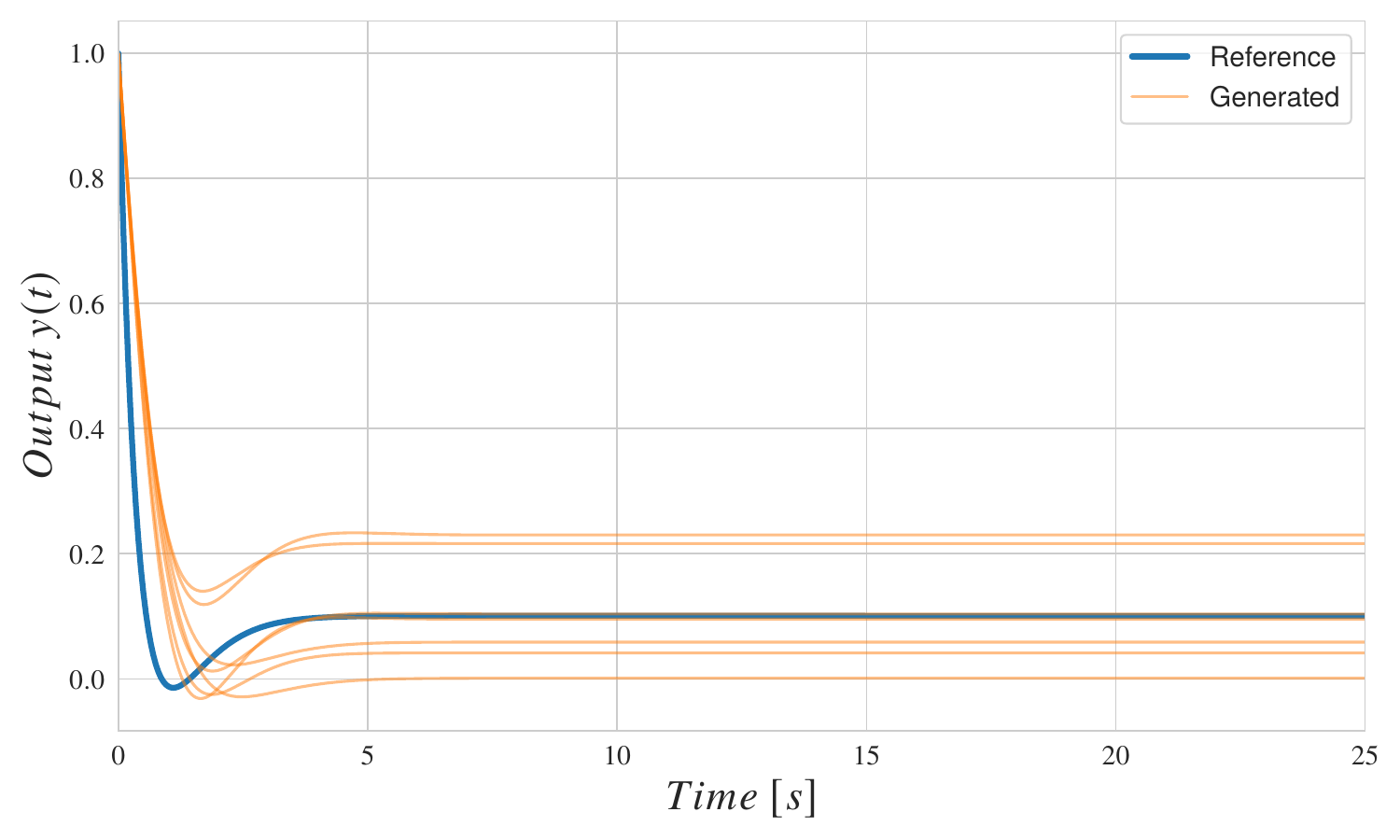}
        \caption{Disturbance rejection.}
    \end{subfigure}
    \caption{Comparison between a reference controller and the controllers generated by the proposed framework for one representative plant.}
    \label{fig:step_and_disturbance}
\end{figure*}

\begin{rem}[Unrealizable or conflicting specifications]
The training set only contains controllers satisfying mild bounds on the metrics (see again Fig.~\ref{fig:dataset_J_dist}). When the user queries the model with targets outside this region, or when specifications are mutually incompatible (\textit{e.g.}, very low settling time together with $\|S(s)\|_{\infty}\approx 1$), the model \textit{cannot satisfy all objectives simultaneously}.  
However, the generated controllers remain close to the feasible boundary, and variance across samples provides a natural diagnostic of specification difficulty.
\end{rem}

We further test the model on a subset of challenging plants satisfying the high-performance targets $\|S(s)\|_{\infty}<1.2$ and settling time $<5\,\mathrm{s}$.  
Figure~\ref{fig:samples_jhat} illustrates that the mean and variance of the generated controllers closely cluster around the targets, highlighting the model’s ability to meet demanding specifications.

Finally, Figure~\ref{fig:step_and_disturbance} compares the time-domain responses of the generated controllers against a reference controller for a representative plant. The generated controllers track the reference response closely in both step and disturbance rejection experiments, confirming that good frequency-domain performance translates into high-quality time-domain behavior.

%%%%%%%%%%%%%%%%%%%%%%%%%%%%%%%%%%%%%%%%%%%%%%%%%%%%%%%%%%%%%%%%%%

\section{Final Discussion}

This work presented a diffusion-based framework for controller synthesis grounded in the Youla--Kučera parameterization. By training a conditional generative model on synthetically generated stabilizing plant--controller pairs, we showed that diffusion models can learn to produce feasible Youla parameters that yield internally stabilizing controllers with user-specified closed-loop performance. The experimental results demonstrate high success rates across a broad variety of unseen plants, providing, to the best of our knowledge, \textit{the first systematic evidence} that diffusion-based generative modeling can be successfully adapted to structured control-design problems while preserving theoretical stability guarantees.

Beyond demonstrating feasibility, our results highlight the potential of diffusion models as a flexible complement to classical controller-synthesis methods. The ability to draw multiple, diverse controllers for the same performance specification enables novel modes of design exploration, uncertainty quantification, and trade-off analysis that are not naturally supported by standard optimization-based techniques. While our experiments focused on second-order SISO plants and two performance metrics, \textit{the framework is general} and can accommodate richer sets of specifications and higher-dimensional systems. Moreover, although we employed classifier-free guidance in this study, other guidance or projection mechanisms can be incorporated with minimal modifications.

From the standpoint of traditional controller design, the proposed approach should be viewed as \emph{complementary} rather than as a replacement for optimization-based methods such as $\mathcal{H}_\infty$ or mixed-sensitivity designs. Classical methods remain highly effective when a single performance objective is specified and an optimization problem is solved once. In contrast, the diffusion-based framework amortizes the design cost across many specifications: once trained, the model can rapidly generate multiple controllers for different performance targets through inexpensive reverse-diffusion passes. This capability is particularly attractive when exploring transient–robustness trade-offs, adapting the controller to varying operating conditions, or handling heterogeneous and user-friendly performance specifications without resorting to weighting-function tuning. We do \textit{not} claim superiority in \textit{per-instance} optimality; rather, the method provides a generative interface to the space of stabilizing controllers, offering a new degree of flexibility in the early stages of control design. A quantitative comparison with LMI-based $\mathcal{H}_\infty$ methods, especially under multiple and potentially conflicting specifications, is an interesting direction for future study.

Looking ahead, several other extensions merit further investigation. Generalizing the framework to higher-order controllers and MIMO nonlinear systems would significantly broaden its applicability. Conditioning on richer closed-loop metrics, such as tracking accuracy, robustness margins, disturbance attenuation, or noise amplification, would enable more expressive and practically relevant design queries. Together, these extensions would further enhance the scalability, interpretability, and practical relevance of diffusion-based controller synthesis.

\bibliography{ifacconf}                          

\begin{thebibliography}{21}
\providecommand{\natexlab}[1]{#1}
\providecommand{\url}[1]{\texttt{#1}}
\providecommand{\urlprefix}{URL }
\expandafter\ifx\csname urlstyle\endcsname\relax
  \providecommand{\doi}[1]{doi:\discretionary{}{}{}#1}\else
  \providecommand{\doi}{doi:\discretionary{}{}{}\begingroup \urlstyle{rm}\Url}\fi

\bibitem[{Bhattacharyya and Keel(2022)}]{bhattacharyya2022linear}
Bhattacharyya, S.P. and Keel, L.H. (2022).
\newblock \emph{Linear multivariable control systems}.
\newblock Cambridge University Press.

\bibitem[{Borase et~al.(2021)Borase, Maghade, Sondkar, and Pawar}]{borase2021review}
Borase, R.P., Maghade, D., Sondkar, S., and Pawar, S. (2021).
\newblock A review of pid control, tuning methods and applications.
\newblock \emph{International Journal of Dynamics and Control}, 9(2), 818--827.

\bibitem[{Chung et~al.(2024)Chung, Kim, Mccann, Klasky, and Ye}]{chung2024diffusionposteriorsamplinggeneral}
Chung, H., Kim, J., Mccann, M.T., Klasky, M.L., and Ye, J.C. (2024).
\newblock Diffusion posterior sampling for general noisy inverse problems.
\newblock \urlprefix\url{https://arxiv.org/abs/2209.14687}.

\bibitem[{De~Bortoli et~al.(2022)De~Bortoli, Mathieu, Hutchinson, Thornton, Teh, and Doucet}]{de2022riemannian}
De~Bortoli, V., Mathieu, E., Hutchinson, M., Thornton, J., Teh, Y.W., and Doucet, A. (2022).
\newblock Riemannian score-based generative modelling.
\newblock \emph{Advances in neural information processing systems}, 35, 2406--2422.

\bibitem[{Dhariwal and Nichol(2021)}]{dhariwal2021diffusionmodelsbeatgans}
Dhariwal, P. and Nichol, A. (2021).
\newblock Diffusion models beat gans on image synthesis.
\newblock \urlprefix\url{https://arxiv.org/abs/2105.05233}.

\bibitem[{Francis(1987)}]{francis1987course}
Francis, B.A. (1987).
\newblock \emph{A course in $H_{\infty}$ control theory}.
\newblock Springer.

\bibitem[{Gahinet and Apkarian(1994)}]{gahinet1994linear}
Gahinet, P. and Apkarian, P. (1994).
\newblock A linear matrix inequality approach to $h_{\infty}$ control.
\newblock \emph{International journal of robust and nonlinear control}, 4(4), 421--448.

\bibitem[{Ho et~al.(2020)Ho, Jain, and Abbeel}]{ho2020denoising}
Ho, J., Jain, A., and Abbeel, P. (2020).
\newblock Denoising diffusion probabilistic models.
\newblock \emph{Advances in neural information processing systems}, 33, 6840--6851.

\bibitem[{Ho and Salimans(2022)}]{ho2022classifierfreediffusionguidance}
Ho, J. and Salimans, T. (2022).
\newblock Classifier-free diffusion guidance.
\newblock \urlprefix\url{https://arxiv.org/abs/2207.12598}.

\bibitem[{Janner et~al.(2022)Janner, Du, Tenenbaum, and Levine}]{janner2022planningdiffusionflexiblebehavior}
Janner, M., Du, Y., Tenenbaum, J.B., and Levine, S. (2022).
\newblock Planning with diffusion for flexible behavior synthesis.
\newblock \urlprefix\url{https://arxiv.org/abs/2205.09991}.

\bibitem[{Karras et~al.(2022)Karras, Aittala, Aila, and Laine}]{karras2022elucidating}
Karras, T., Aittala, M., Aila, T., and Laine, S. (2022).
\newblock Elucidating the design space of diffusion-based generative models.
\newblock \emph{Advances in neural information processing systems}, 35, 26565--26577.

\bibitem[{Khalil et~al.(1996)Khalil, Doyle, and Glover}]{khalil1996robust}
Khalil, I., Doyle, J., and Glover, K. (1996).
\newblock \emph{Robust and optimal control}, volume~2.
\newblock Prentice hall New York.

\bibitem[{Lee et~al.(2023)Lee, Lu, and Tan}]{lee2023convergence}
Lee, H., Lu, J., and Tan, Y. (2023).
\newblock Convergence of score-based generative modeling for general data distributions.
\newblock In \emph{International Conference on Algorithmic Learning Theory}, 946--985. PMLR.

\bibitem[{Mahtout et~al.(2020)Mahtout, Navas, Milanes, and Nashashibi}]{Mahtout2020Advances}
Mahtout, I., Navas, F., Milanes, V., and Nashashibi, F. (2020).
\newblock Advances in youla-kucera parametrization: A review.
\newblock \emph{Annual Reviews in Control}, 49, 81--94.
\newblock \doi{https://doi.org/10.1016/j.arcontrol.2020.04.015}.

\bibitem[{Oko et~al.(2023)Oko, Akiyama, and Suzuki}]{oko2023diffusion}
Oko, K., Akiyama, S., and Suzuki, T. (2023).
\newblock Diffusion models are minimax optimal distribution estimators.
\newblock In \emph{International Conference on Machine Learning}, 26517--26582. PMLR.

\bibitem[{Skogestad and Postlethwaite(2005)}]{skogestad2005multivariable}
Skogestad, S. and Postlethwaite, I. (2005).
\newblock \emph{Multivariable feedback control: analysis and design}.
\newblock john Wiley \& sons.

\bibitem[{Sohl-Dickstein et~al.(2015)Sohl-Dickstein, Weiss, Maheswaranathan, and Ganguli}]{sohldickstein2015deepunsupervisedlearningusing}
Sohl-Dickstein, J., Weiss, E.A., Maheswaranathan, N., and Ganguli, S. (2015).
\newblock Deep unsupervised learning using nonequilibrium thermodynamics.
\newblock \urlprefix\url{https://arxiv.org/abs/1503.03585}.

\bibitem[{Song et~al.(2021)Song, Sohl-Dickstein, Kingma, Kumar, Ermon, and Poole}]{song2021scorebased}
Song, Y., Sohl-Dickstein, J., Kingma, D.P., Kumar, A., Ermon, S., and Poole, B. (2021).
\newblock Score-based generative modeling through stochastic differential equations.
\newblock In \emph{International Conference on Learning Representations}.
\newblock \urlprefix\url{https://openreview.net/forum?id=PxTIG12RRHS}.

\bibitem[{Youla et~al.(1976)Youla, Bongiorno, and Jabr}]{Youla1976Optimal}
Youla, D., Bongiorno, J., and Jabr, H. (1976).
\newblock Modern wiener--hopf design of optimal controllers part i: The single-input-output case.
\newblock \emph{IEEE Transactions on Automatic Control}, 21(1), 3--13.
\newblock \doi{10.1109/TAC.1976.1101139}.

\bibitem[{Zhou and Doyle(1998)}]{zhou1998essentials}
Zhou, K. and Doyle, J.C. (1998).
\newblock \emph{Essentials of robust control}, volume 104.
\newblock Prentice hall Upper Saddle River, NJ.

\bibitem[{Zirvi et~al.(2025)Zirvi, Tolooshams, and Anandkumar}]{zirvi2025diffusionstateguidedprojectedgradient}
Zirvi, R., Tolooshams, B., and Anandkumar, A. (2025).
\newblock Diffusion state-guided projected gradient for inverse problems.
\newblock \urlprefix\url{https://arxiv.org/abs/2410.03463}.

\end{thebibliography}
\end{document}